\documentclass[12pt]{article}
\usepackage{graphicx}
\usepackage{amssymb,amsmath,amsfonts,palatino,amsthm}
\usepackage{amssymb}
\usepackage{epstopdf}
\newcommand{\beq}{\begin{equation}}

\newcommand{\be}{\begin{eqnarray}}

\newcommand{\ee}{\end{equation}}

\def\Z#1{_{\lower2pt\hbox{$\scriptstyle#1$}}}

\begin{document}

\title{We Probably Live On An Inflating Brane-World \footnote{(Slightly expanded version of) essay written for the
Gravity Research Foundation 2011 essay contest.}}
\author{Ishwaree P. Neupane\footnote{E-mail: ishwaree.neupane@canterbury.ac.nz}
\\
Department of Physics and Astronomy, University of Canterbury\\
Private Bag 4800, 8041 Christchurch, New Zealand}
\date{31 March 2011}
\maketitle

\begin{abstract}

\noindent

Brane-world models where observers are trapped within the
thickness of a 3-brane offer novel perspectives on gravitation and
cosmology. In this essay, I would argue that the problem of a late epoch
acceleration of the universe is well explained in the framework of a
4-dimensional de Sitter universe embedded in a 5-dimensional de
Sitter spacetime. While a 5D anti de Sitter space background is
important for studying conformal field theories -- for
its role in the AdS/CFT correspondence -- the existence of a 5-dimensional
de Sitter space is crucial for finding an
effective 4D Newton constant that remains finite and a
normalizable zero-mode graviton wave function.

\end{abstract}
\newpage


Cosmology is a young science -- one which attempts to reconstruct
and explain the entire history of the universe from nearly 13.8
billions of years ago. This cosmic history should include a brief
period of rapid exponential expansion of the early universe, known
as `inflation', which provides the most credible explanation to
how causal physics in a very early epoch produced the large scale
structures that we observe today. However, looking back so far in
time (via observations) is extremely difficult and an added
difficulty is that many of the theoretical pillars of physics upon
which the models of inflation and late epoch cosmic acceleration
(attributed to dark energy) rest have only been proposed within
the last 3 decades or so. That hasn't given cosmologists (and
astrophysicists) much time to fully flesh out and comprehend the
situation. There are two distinct possibilities: either we're
missing new physics and hence going to find a fatal flaw in our
prevailing view of the universe's present composition, or that we
encounter interesting discoveries and surprises in cosmology and
particle physics experiments waiting ahead of us.


It is true that our understanding of the physical universe has
deepened profoundly in the last few decades through thoughts,
observations and experiments. It is also true that the concurrent
universe still has a number of cosmological mysteries, but the one
that has most puzzled physicists is the smallness of the present
value of gravitational vacuum energy (or `dark energy'), i.e.
$\Lambda_{\rm obs}\lesssim (10^{-3}~{\rm eV})^4$ and its eminent
effect on an accelerated expansion of the universe at a late
epoch~\cite{supernovae}. There is no shortage of ideas for how to
construct a model which is capable to produce a universe with de
Sitter-type expansion. There is a large gamut of gravitational
theories (see, e.g.~\cite{Copeland06,Ish-Review2007}) that are
capable to explain an accelerated expansion of the universe with
certain modification of Einstein's theory of general relativity.
In this essay, I would argue that the problem of a late epoch
acceleration as well as the smallness of the cosmological constant
may be well explained within the framework of a de Sitter 3-brane
embedded in a 5D de Sitter spacetime (dS$_5$).

The issue of a late epoch cosmic acceleration (or the dark energy
problem) is perhaps not about difficulties of finding a particular
cosmological model which could mimic as the Lambda-CDM cosmology,
described by Einstein gravity with a cosmological constant and
minimally coupled to both the luminous (baryonic) and non-luminous
(cold dark) matter. The challenge is to come up with a fully
consistent theory in four-dimensions that explains the origin of
cosmic acceleration, while providing insights into some other
major problems in physics, including the mass hierarchy problem in
particle physics. More generally, we should be able to explain why
the constants of the standard model of cosmology, including the
dark energy and dark matter densities, have the values they do.

The paradigm that the physical universe is a brane-like
4-dimensional hypersurface embedded in a higher dimensional space
is fascinating~\cite{Domain-wall,RS2}. This very idea is inspired
by a theoretical framework of scientific thoughts and also by
fundamental theories of gravity, particles and fields, such as
string and M theory, which provide novel approaches for unifying
Einstein's general relativity with quantum field theories.
Obtaining 4-dimensional de Sitter solutions from dynamical
compactifications of higher dimensional theories has been an
important issue in string cosmology. Since an epoch of cosmic
acceleration plays an important role in modern cosmological
models, it would be very interesting to know whether this effect
can be explained within the framework of string theory and/or
modified theories of gravity, such as, brane-world gravity in
higher dimensions~\cite{Ish-0910,Ish09b}.

An attractive feature of 5D brane-world models is that the
standard 4-dimensional gravity, not very different from Einstein's
theory, is realized as the zero-mode solution of a 5D graviton
wave equation. In the simplest RS brane-world proposal~\cite{RS2},
the background spacetime geometry is a 5D Anti de Sitter space
(AdS$_5$), which is warped. The all 4 spatial dimensions are
noncompact, but one of them behaves differently for particles and
fields which are confined within the brane's thickness. There also
exists a massless graviton as the zero-mode solution, which
reproduces the standard Newtonian gravity on the
3-brane~\cite{Roy,Brevik-etal,Kehagias02,Sasaki2000a}. The
Kaluza-Klein modes arising as the effect of graviton fluctuations
in the bulk spacetime give rise to corrections to the Newton's
force law~\cite{Garriga,Ghoroku,2010prd}.

The RS brane-world model~\cite{RS2} corresponds to a static
universe for which the Hubble expansion parameter is zero. This
original setup is not suitable for describing a realistic
cosmology for which space and time are not independent (or the
3-brane is dynamical). Furthermore, AdS$_5$ is perhaps {\it not}
the most preferred background geometry for a universe to
experience a de Sitter-type expansion at a late epoch. The main
reason being that the standard 4D gravity, which may be viewed as
the zero-mode solution of a 5D graviton wave equation, is not
necessarily normalizable if the background geometry is AdS$_5$.

To quantify this, one may consider a 5-dimensional metric of the following form %
\begin{eqnarray} ds^2= e^{2A(z)} \left(\gamma_{\mu\nu} dx^\mu dx^\nu + dz^2
\right), \end{eqnarray} where $e^{2A(z)}$ is the warp factor, and
look for a class of solutions for which the 4D line-element is the
standard Friedmann-Lama\^itre-Robertson-Walker (FLRW) metric,
\begin{eqnarray}\label{FRW}
ds_4^2 &\equiv&  \gamma_{\mu\nu} dx^\mu dx^\nu =  - dt^2+
a^2(t)\left[\frac{ dr^2}{1-k r^2}+ r^2 d\Omega_2^2\right],
\end{eqnarray}
where $k$ is the 3D curvature constant with the dimension of
inverse length squared. The relevant 5D Einstein-Hilbert
(gravitational) action is
\begin{equation}\label{5d-gravi}
S_{\rm grav}  = M_{(5)}^3 \int d^5{x} \sqrt{-g}
\left(R-2\Lambda_5\right) + \int \sqrt{-\gamma}\,(-\tau),
\end{equation}
where $M_{(5)}$ is the 5D Planck mass and $\tau$ is the 3-brane
tension. If the size of the physical three dimensions does not
change with time or $ds_4^2= \eta_{\mu\nu}dx^\mu dx^\nu \equiv
-dt^2 + dx^2 +dy^2 + dz^2$, then the zero-mode gravity solution is
normalizable only when (i) the bulk cosmological constant is
negative, i.e. $\Lambda_5 \equiv - 6/\ell^2$, where $\ell$ is the
radius of curvature of the 5D bulk spacetime, and (ii) the 3-brane
tension satisfies the relation $\tau= 12 M_{(5)}^3/\ell$.

In the limit $\Lambda_5\to 0$, the 5D bulk spacetime is spatially
flat and gravity is {\it not} localized. In this case the solution
is trivial, i.e. $A(z)=0$, $a(t)=1$. However, when
$\Lambda_5=6/\ell^2
>0$ and $a(t)\propto e^{Ht}$, the solution becomes
nontrivial, which is given by
\begin{equation}
e^{A(z)} =\frac{\ell H}{\cosh H z}, \qquad \tau=\frac{6
M_{(5)}^3}{\ell}\sinh H z_c,\label{flat-brane-T}
\end{equation}
where $z_c$ ($>0$) is a constant. Upon the dimensional reduction
from 5D to 4D, we find that the 4D effective Planck mass $M_{\rm
{Pl}}$ is related to the 5D Planck mass $M_{(5)}$ via
\begin{equation}\label{5to4Planck}
M_{\rm Pl}^2= {\pi \ell^3 H^2\over 2}\,M_{(5)}^3.
\end{equation}
Note that unlike in the RS static brane-world models, for which
$\ell_{AdS}\lesssim 0.01 ~{\rm mm}$, the de Sitter curvature
radius $\ell$ can be significantly larger; if we want $M_{(5)} >
10^{-15}\,M_{\rm Pl}$ and $H= H_{\rm obs} \sim 10^{-61}\,M_{\rm
Pl}$, then $\ell\lesssim  10^{55}\,\ell_{\rm Pl}\sim 10^{22}\,{\rm
cm}$. The 4D Planck mass is well defined even in the limit $H\to
0$ since when the thickness of the brane becomes infinitely large
(or $H^{-1}\to \infty$), $\ell$ can take a naturally large value.
In such a case the background 5D geometry is only modestly warped
so as to balance the effect of a small positive curvature
associated with 4D cosmological constant on the de Sitter 3-brane
and $\Lambda_4= 6H^2$~\cite{Ish09a}.

Let us now consider the linear perturbations of the 5D metric
$\delta^{(5)}g_{AB} = h_{AB}$. For $\Lambda_5>0$, we find that the
transverse-traceless tensor modes $\delta g_{ij}= h_{ij}(x^\mu,
z)\equiv \delta_i^\mu \delta_j^\nu h_{\mu\nu}(x^\mu, z)= \sum
\alpha_m(t) \,e^{-3A/2} \psi_m e^{i k\cdot x} \hat{e}_{ij}$
satisfy the following Schr\"odinger-type wave equation
\begin{equation}
\frac{d^2\psi_m}{dz^2} - V(z) \psi_m = - m^2 \psi_m,
\label{Schro-main}
\end{equation}
where
\begin{equation}
V(z) = \frac{9H^2}{4}- \frac{15 H^2}{4 \cosh^2 (Hz)}-6 H
\tanh(Hz)\delta(z-z_c).\label{PositiveV}
\end{equation}
The brane is located at $z=z_c$. The zero-mode solution ($m^2=0$)
is given by
\begin{equation}
\psi\Z{0}(z)=\frac{b\Z{0}}{(\cosh(H z))^{3/2}},\label{sol1}
\end{equation}
which is clearly normalizable since
$$ \int_{-\infty}^{\infty} |\psi_0(z)|^2 dz= \frac{\pi b\Z{0}^2}{2H}.$$
There is one more bound state solution, i.e.,
\begin{equation}
\psi\Z{1}(z) \propto \frac{\sqrt{\cosh^2(H z)-1}}{(\cosh(H
z))^{3/2}},\label{sol2}
\end{equation}
which is obtained by taking $m^2=2H^2$. This solution is also
normalizable. However, only the zero-mode solution ($m^2=0$) is
localized on the de Sitter 3-brane~\cite{2010prd}.

Around the brane's position at $z\equiv z_c $,
satisfying $H z \ll 1$, the solution looks like
\begin{equation}
\psi(z) = c\Z{1} P_\mu + c\Z{2} Q_\mu,
\end{equation}
where
\begin{equation}
P_\mu = 1- \frac{3+2\zeta}{4}(H z)^2 + \cdots,\quad Q_\mu = H z
-\frac{3+2\zeta}{12} (H z)^3 +
 \cdots,
\end{equation}
and $\zeta\equiv m^2/H^2\ge 9/4$. This solution corresponds to a
situation that the thickness of the brane is much larger than the
size of the fifth dimension, $H^{-1} \gg z$. In the large $z H$
limit,
\begin{equation}
\psi\Z{Hz\to \infty} =c_1\, e^{i\mu z H} + c_2\, e^{-i\mu z H},
\end{equation}
where $\mu\equiv \sqrt{\zeta-\frac{9}{4}}$. The masses of
Kaluza-Klein modes are quantized in units of $H$. With $c\Z{1}=0$,
and away from the brane's position, satisfying $z\gg H^{-1}$, all
heavy modes with $\mu> 0$ become oscillating plane waves, which
represent the de-localized KK modes.

To estimate the correction to Newton's force law generated by a
discrete tower of Kaluza-Klein modes, one may go to the thin brane
limit, i.e. $H^{-1} \to 0$, but keeping the ratio $z_c/H^{-1}$
finite. One would also assume that the matter fields in the
4-dimensional theory is smeared over the width of the brane and
the brane thickness is smaller compared with the bulk curvature,
$H^{-1} < \ell$, so $H\ell > 1$. Under these approximations, the
gravitational potential between two point-like sources of masses
$M_1$ and $M_2$ located on the brane is modified via exchange of
gravitons living in 5 dimensions as
\begin{eqnarray}
U(r)&\simeq & \frac{G_4 M_1 M_2}{r} \left(1+ {2\alpha\over \pi}
{M_{\rm Pl}^2\over M_{(5)}^3 r} \sum_i e^{-m_i r}  \right),
\end{eqnarray}
where $m_i \ge \sqrt{2} H$, $r$ is the distance between the two
pointlike sources, and $\alpha$ is a constant of order unity. The
correction to the gravitational potential due to the massive KK
states may dominate for $r < H^{-1}< \ell $, leading to a
5-dimensional behavior which is not Newtonian. However, since the
corrections to the gravitational potential are suppressed by a
factor of $\sum_i e^{-m_i r}$, with $m_i \gtrsim {\rm TeV} \sim
10^{-15}\,{\rm cm}$, the deviation from Newton's inverse law may
not show up unless we probe a sufficiently small distance scale,
$r\lesssim 10^{-12}~{\rm cm}$.

For $\Lambda_5 < 0$, the analysis closely follows the one given
above, but the opposite sign of $\Lambda_5$ modifies the above
results. The Schr\"odinger-type potential $V(z)$ in
(\ref{Schro-main}) is given by
\begin{equation}
V(z) = \frac{9H^2}{4}+ \frac{15 H^2}{4 \sinh^2 (Hz)}-6 H
\coth(Hz)\delta(z-z_c).
\end{equation}
The zero-mode solution ($m^2=0$) is now given by
\begin{equation}
\psi\Z{0}(z)=\frac{c_0}{(\sinh(H z))^{3/2}}. \label{sol3}
\end{equation}
In this case it is necessary to have ${\mathbb Z}_2$-symmetry, or
allow a cutoff scale, $0< z_c \le z$; otherwise the zero-mode
solution is non-normalizable.

A similar analysis can be carried out by introducing scalar
fields~\cite{DeWolfe99}. Here we study a 5D gravity action
minimally coupled with a canonical scalar field $\phi$ with
potential $U(\phi)$:
\begin{equation} S_{\rm grav}={1\over 2} \int d^5{x} \sqrt{-g}
\left[\frac{R}{\kappa_5^2} - g^{AB} \left(\partial_A \phi\right)
\left(\partial_B \phi\right) - 2 U(\phi) \right],
\end{equation}
where $\kappa_5^2 \equiv 1/M_{(5)}^3$. For simplicity, we begin
with the $k=0$ (spatially flat) case. It is natural to assume that
$\phi$ depends only on $z$. The 5D field equations now take the
form
\begin{eqnarray}
{\phi^\prime}^2 &=& \frac{3}{\kappa_5^2}
\left({A^\prime}^2-A^{\prime\prime}-{H^2}\right), \qquad U(\phi) =
\frac{3 \,e^{-2A}}{2 \kappa_5^2} \left({3 H^2} -3 {A^\prime}^2 -
A^{\prime\prime}\right),\label{eq2}
\end{eqnarray}
where ${}^\prime \equiv d/dz$. These equations admit a series of
solutions with different choice of $\phi(z)$ or $A(z)$. For
illustration, we look for a particular solution of the form
\begin{equation}
A(z)= \beta- \lambda \ln \cosh \left( \frac{H
z}{\lambda}\right),\label{sol-Az}
\end{equation}
where $\beta$ and $\lambda$ are some constants. This leads to a
standard domain-wall-type solution
\begin{equation}
\kappa_5 \phi= \phi\Z0 - \phi\Z{1} \arcsin \tanh \left( \frac{
Hz}{\lambda}\right),\label{scalar-anz}
\end{equation}
where $\phi_i$ are dimensionless constants, and $\phi\Z{1}=
\sqrt{3\lambda(1-\lambda)}$. Note that $\phi$ behaves as a
canonical scalar field when $0<\lambda<1$. The scalar potential
$U(\phi)$ is now given by
\begin{equation}
U(\phi)=\frac{3(1+3\lambda) H^2}{2\lambda \kappa_5^2}
\left[\cos^2\left(\frac{\phi\Z{0}-\kappa_5
\phi}{\phi\Z{1}}\right)\right]^{(1-\lambda)}.\label{sol-5d-V}
\end{equation}
This potential is of Mexican-hat type (as long as $0<\lambda<1$)
and the maximum occurs at $z=0 $. In fact, the solutions
(\ref{sol-Az})-(\ref{sol-5d-V}) are available also when the 3D
spatial curvature $k$ is nonzero,
 i.e. $a(t)={a_0\over 2} \,e^{Ht} + {k\over 2 H^2 a_0}\,
e^{-Ht}$. For $k\ne 0$, the choice $H=0$ is {\it not} physical.

To obtain an effective 4D theory, we shall consider the
dimensionally reduced action
\begin{eqnarray}
S_{\rm eff}  &=& {M_{\rm Pl}^2\over 2} \int d^4{x} \sqrt{-g_4}
\left( {R}_4 - 2 \Lambda\Z{4}\right),
\end{eqnarray}
where
\begin{eqnarray}
M_{\rm Pl}^2 = e^{3\beta}\,{M_{(5)}^3 \lambda\over  H} \int
\frac{d\varphi}{\left( \cosh{\varphi}\right)^{3\lambda}}, \qquad
\Lambda\Z{4} = e^{3\beta}\,{M_{(5)}^3 H\over M_{\rm Pl}^2} \int
{\left(6\lambda \cosh^2\varphi-3\lambda -1\right)\over
{(\cosh\varphi)}^{2+3\lambda}} \,d\varphi
\end{eqnarray}
and $\varphi\equiv (H z/ \lambda) $. Note that the integral
\begin{equation}
I\Z{\lambda}\equiv \int_{-\infty}^{\infty}
\frac{d\varphi}{\left(\cosh{\varphi}\right)^{3\lambda}}=\frac{\sqrt{\pi}\,
\Gamma\left[3\lambda/2\right]}{\Gamma\left[(3\lambda+1)/2\right]}
\end{equation}
is finite with $\lambda>0$. This shows that the 4D effective
Planck mass $M_{\rm Pl}$ is also finite. That is, as in the model
without a scalar field, the effective 4D Newton's constant is
finite despite having a noncompact extra dimension.
%
One may extend the above discussion for a general class of
$D$-dimensional warped supergravity models, including
10-dimensional supergravity coupled with scalar fields and form
field strengths. It is possible to obtain a 4-dimensional de
Sitter universe as the exact solution of classical supergravities
in general $D=4+n$ dimensions~\cite{Ish10d}. It is also possible
to stabilise the scale (or size) of the internal space by suitably
choosing fluxes (i.e. the gauge field strengths wrapped around the
internal manifold), the curvature term associated with the
internal space and the bulk cosmological constant, leading to a
dynamical mechanism of warped compactification.

\medskip
I conclude with a few remarks.

One of the most exciting developments in cosmology is the
suggestion that the warping of extra dimensions plays a key role
in explaining mass hierarchy and localization of gravity on a de
Sitter 3-brane, embedded in a higher dimensional spacetime. dS$_5$
space allows a foliation by a flat space but that is a spacelike
hypersurface. There is no way to cut dS$_5$ by Minkowski
spacetime. That is, in a cosmological setting, a 4D Minkowski
spacetime (or a flat 3-brane) is not a solution to 5D Einstein
equations, if the bulk spacetime geometry is de Sitter. This is in
contrast to the results in Randall-Sundrum brane-world models in
AdS$_5$ spacetimes. But this is anyway not a problem since the
universe has probably never gone through a phase of being close to
a static universe or a flat 3-brane.

While a 5D anti de Sitter space background is important for
studying conformal field theories -- for its role in the AdS/CFT
correspondence -- the existence of a 5-dimensional de Sitter space
is crucial for finding an effective 4D Newton constant that
remains finite and a normalizable zero-mode graviton wave
function. The 5-dimensional model discussed above is found to
admit both an effective 4-dimensional Newton constant that remains
finite and a normalizable graviton wave function.

In this essay, we tried to explore a much desired connection
between 5D de Sitter gravity and FRW-type cosmologies built upon
the idea of the existence of an extra noncompact dimension and a
warped background geometry in five dimensions. The important
question that we seek to address is whether or not a canonical
theory of brane-world gravity admits a lower dimensional
description in which gravity emerges from it in what would now be
called a consistent warped compactification to 4D de Sitter
spacetime.

It may well be that the final theory of quantum gravity requires a
highly symmetric (or even supersymmetric) background geometry in
1+9 spacetime dimensions, such as ${\rm AdS}_5 \times S^5$ - a
product space of a 5-dimensional Anti de Sitter space and a
five-sphere. But such geometry wouldn't remain as a stable
background once the universe starts to expand or undergoes an
inflationary de Sitter phase. Nevertheless, it's possible to
obtain a realistic cosmology in four-dimensions by allowing
spacetime backgrounds such as dS$_5\times S^5$ and dS$_5 \times
T^{1,1}$. This outcome possibly implies that a spontaneous
breaking of supersymmetry corresponds to a tunnelling of the
universe from an AdS$_5$ to a dS$_5$ state.


\section*{Acknowledgement}

This work was supported by the Marsden fund of the Royal Society
of New Zealand.



\begin{thebibliography}{99}
\itemsep 0pt

\bibitem{supernovae}
A. G. Riess {\it et al.}
  Astron.\ J.\  {\bf 116} (1998) 1009;
  S. Perlmutter {\it et al.}, Astrophys. J. {\bf 517} (1999) 565.

\bibitem{Copeland06}
  E.~J.~Copeland, M.~Sami, S.~Tsujikawa,
  Int.\ J.\ Mod.\ Phys.\  {\bf D15}, 1753-1936 (2006).

\bibitem{Ish-Review2007}
  I.~P.~Neupane,
  arXiv:0711.3234 [hep-th].


\bibitem{Domain-wall}
  V.~A.~Rubakov and M.~E.~Shaposhnikov,
  Phys.\ Lett.\  B {\bf 125} (1983) 136;\\
%
M.~Visser,
  Phys.\ Lett.\  {\bf B159} (1985) 22.


\bibitem{RS2}
  L.~Randall and R.~Sundrum,
  Phys.\ Rev.\ Lett.\  {\bf 83} (1999) 4690.


\bibitem{Ish-0910}
I.~P.~Neupane,
  Class.\ Quant.\ Grav.\  {\bf 26} (2009) 195008.\\
%
  I.~P.~Neupane,
  Class.\ Quant.\ Grav.\  {\bf 27} (2010) 045011.

\bibitem{Ish09b}
  I.~P.~Neupane,
  Phys.\ Lett.\  B {\bf 683} (2010) 88. 

\bibitem{Roy} R.~Maartens,
  Living Rev.\ Rel.\  {\bf 7} (2004) 7.

\bibitem{Brevik-etal}
  I.~H.~Brevik, K.~Ghoroku, S.~D.~Odintsov and M.~Yahiro,
  Phys.\ Rev.\  D {\bf 66} (2002) 064016.

\bibitem{Kehagias02}
  A.~Kehagias and K.~Tamvakis,
  Class.\ Quant.\ Grav.\  {\bf 19} (2002) L185.


\bibitem{Sasaki2000a}
  U.~Gen and M.~Sasaki,
  Prog.\ Theor.\ Phys.\  {\bf 105} (2001) 591.

\bibitem{Garriga} J.~Garriga and T.~Tanaka,
  Phys.\ Rev.\ Lett.\  {\bf 84} (2000) 2778.

\bibitem{Ghoroku}
  K.~Ghoroku, A.~Nakamura and M.~Yahiro,
  Phys.\ Lett.\  B {\bf 571} (2003) 223.

\bibitem{2010prd}
  I.~P.~Neupane,
  Phys.\ Rev.\  {\bf D83} (2011) 086004.

\bibitem{Ish09a}
I.~P.~Neupane, 
Int. J. Mod. Phys. D {\bf 19} (2010) 2281.

\bibitem{DeWolfe99}
  O.~DeWolfe, D.~Z.~Freedman, S.~S.~Gubser and A.~Karch,
  Phys.\ Rev.\  D {\bf 62} (2000) 046008.



\bibitem{Ish10d} I.P. Neupane,
Nucl.\ Phys.\  {\bf B847} (2011) 549 


\end{thebibliography}
\end{document}